\documentclass[pre, aps, longbibliography, twocolumn, reprint, showpacs, nofootinbib, superscriptaddress]{revtex4-1} 
\usepackage{graphicx}
\usepackage{amssymb}   
\usepackage{amsfonts}
\usepackage{amsmath}
\usepackage{dsfont}
\usepackage{dcolumn}   
\usepackage{bm}        
\usepackage[mathscr]{eucal}
\usepackage[dvipsnames]{xcolor}
\usepackage[colorlinks, linkcolor=DarkOrchid,citecolor=DarkOrchid,urlcolor=PineGreen]{hyperref}
\usepackage[all]{hypcap} 
\usepackage{mathtools}
\usepackage{wasysym}
\usepackage{tikz}

\newcommand{\mc}{\mathcal}

\newcommand{\ra}{\rangle}
\newcommand{\la}{\langle}

\newcommand{\beq}{\begin{equation}}
\newcommand{\eeq}{\end{equation}}
\usepackage[normalem]{ulem}
\newcommand\redsout{\bgroup\markoverwith{\textcolor{red}{\rule[0.5ex]{2pt}{0.4pt}}}\ULon}

\begin{document}

\title{Family-Vicsek universality of the binary intrinsic dimension of nonequilibrium data}

\author{Roberto Verdel}
\email{rverdel@ictp.it}
\affiliation{The Abdus Salam International Centre for Theoretical Physics (ICTP), Strada Costiera 11, 34151 Trieste, Italy}

\author{Devendra Singh Bhakuni}
\affiliation{The Abdus Salam International Centre for Theoretical Physics (ICTP), Strada Costiera 11, 34151 Trieste, Italy}

\author{Santiago Acevedo}
\affiliation{International School for Advanced Studies (SISSA), Via Bonomea 265, 34136 Trieste, Italy}

\date{\today}

\begin{abstract}

%
The intrinsic dimension (ID) is a powerful tool to detect and quantify correlations from data. Recently, it  has been successfully applied to study statistical and many-body systems in equilibrium, yet its application to systems away from equilibrium remains largely unexplored. 
Here we study the ID of nonequilibrium growth dynamics data, and show that even after reducing these data to binary form, their binary intrinsic dimension (BID) retains essential physical information. 
Specifically, we find that, akin to the surface width, it exhibits Family-Vicsek dynamical scaling---a fundamental feature to describe universality in surface roughness phenomena.
These findings highlight the ability of the BID to correctly discern key properties and correlations in nonequilibrium data, and open an avenue for  alternative characterizations of out-of-equilibrium  dynamics.

\end{abstract}

\maketitle


\emph{Introduction}. Identifying the relevant correlations that give rise to the interesting properties of a system is a general task at the core of multiple disciplines. 
In physics, many-particle systems provide a compelling example, where correlations can lead to emergent collective behavior that cannot be explained from the properties of individual constituents alone~\cite{Kardar_2007, Altland_Simons_2010}, 
yet providing complete microscopic descriptions of such systems is one of the most formidable challenges in physics.

Recently, data science tools have been used in the quest of formulating general frameworks to characterize complex systems, in approaches complementary to traditional statistical mechanical treatments.  Among such tools, the \emph{intrinsic dimension} (ID)---a key concept in manifold learning~\cite{annurev-statistics-040522-115238}, stands out as a highly effective means for quantifying correlations.  Specifically, the ID can be interpreted as the least number of degrees of freedom needed to describe a collection of observations (dataset) of a set of many variables~\cite{CAMASTRA201626}. 
Heuristically, the more correlations there are among the variables, the fewer effective degrees of freedom, and thus the lower the ID of the dataset.
This property has driven the current use of the ID to understand information processing and data representations in artificial neural networks~\cite{NEURIPS2019_cfcce062, Gong_2019_CVPR, pope2021the, NEURIPS2023_a0e66093, acevedo2025unsupervised}, and to study phase transitions and critical phenomena in many-body systems at equilibrium~\cite{PhysRevX.11.011040, PRXQuantum.2.030332, PhysRevE.109.034102, 10.21468/SciPostPhysCore.6.4.086, PhysRevB.108.134422, acevedo2025unsupervised}. However, in the physics context, little is known about its applicability to describe data from out-of-equilibrium systems, apart from a few recent efforts~\cite{PhysRevB.106.144313, PhysRevB.109.075152, Cao_2024}.

\begin{figure}[bt!]
	\centering
	\includegraphics[width=\columnwidth]{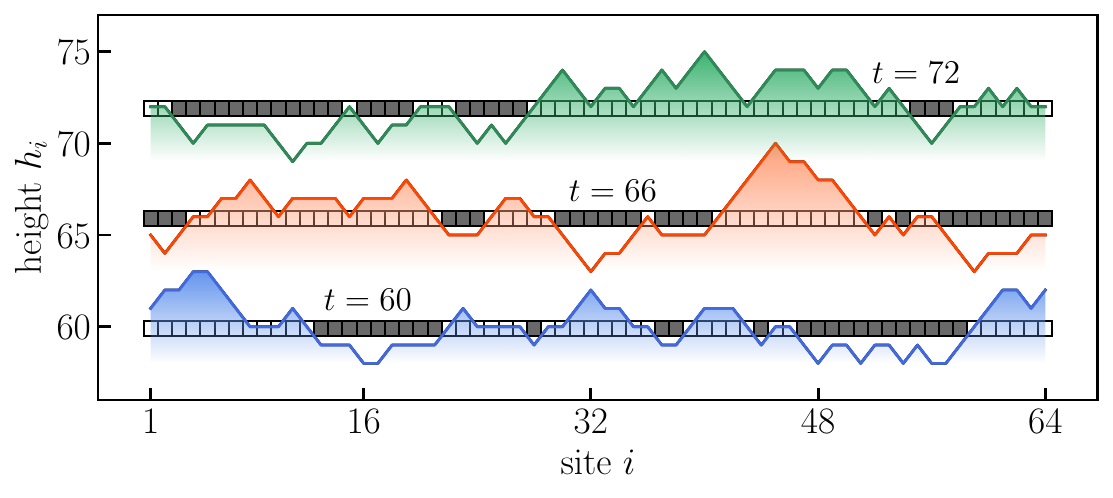}
	\caption{ 
   Typical surface profiles (rough curves) of a nonequilibrium discrete growth process at different times, juxtaposed with their binary representation (rows of light and dark boxes, representing $+1$ and $-1$ values, respectively). Such a binary representation is obtained by checking if the value of each height variable $h_i$ is below or above (or equal to) the average height $\overline{h}$ of the given surface profile [see Eq.~\eqref{eq:binary}].
    }
	\label{fig:snapshots}
\end{figure}

In this work, we investigate the ID of \emph{nonequilibrium data} associated with nonequilibrium growth processes. Due to desirable scaling properties discussed below, we use a recently developed ID estimator for binary data~\cite{acevedo2025unsupervised}, hereafter referred to as BID. To this end, we first reduce the data to binary form;  see Fig.~\ref{fig:snapshots}. 
From a physical perspective, this can be seen as a ``coarse-graining'' procedure, \emph{at the data level}, aimed to remove unnecessary details of the system, in the same spirit of the renormalization group~\cite{PhysicsPhysiqueFizika.2.263, kadanoff2000statistical}. 
Beyond this fundamental perspective, such forms of data compression are also important for applications. For instance, in deep learning, they go under the name of ``quantization methods''~\cite{hubara2018quantized,guo2018survey,courbariaux2016binarized,wang2023bitnet}, and are nowadays standard protocols to save time, memory, and energy when dealing with models consisting of billions of parameters. 
In this context, binarization has shown nice properties in high dimensions, such as approximately preserving the scalar product between activations and weights~\cite{anderson2017high}, as well as data neighborhoods~\cite{acevedo2025unsupervised}.
Here we focus on statistical lattice models for nonequilibrium roughening, which is key to describing a plethora of real-world systems~\cite{HALPINHEALY1995215} and has gained renewed attention very recently, particularly in quantum physics~\cite{PhysRevX.7.031016, PhysRevLett.124.210604, PhysRevLett.127.090601, PhysRevLett.129.110403, Fontaine2022, PhysRevLett.129.230602, PhysRevLett.132.130401, PhysRevB.109.035164,   PhysRevB.110.014203, krinitsin2025roughening, tater2025bipartite, moca2025dynamic}.
Our central observation is that the BID of correlated growth data---even after being highly compressed---exhibits Family-Vicsek dynamical scaling~\cite{Family_1985, PhysRevLett.52.1669}.
The latter notion is important because it provides a framework for understanding universal aspects of surface roughness dynamics.
Thus, our results unveil a profound relation between fundamental concepts of current interest in physics (dynamical scaling) and data science (intrinsic dimension). 
Further, they also demonstrate the BID's effectiveness to correctly quantify correlations after having carried out data compression via binarization.


\emph{The binary intrinsic dimension}.   
Let us consider a binary dataset consisting of $N_r$  realizations of $\mc{N}$ spin variables, that is, $\{\vec{\sigma}^{(s)}\}_{s=1}^{N_r}$, where $\vec{\sigma}^{(s)}\in \{-1, +1\}^\mc{N}$. If such variables are statistically independent, the data points $\vec{\sigma}^{(s)}$ will be uniformly distributed over the vertices of the $\mc{N}$-dimensional hypercube. The ID of such dataset must then be equal to $\mc{N}$, and it is reflected in the distribution of the Hamming distance $r_{H}(\vec{\sigma}^{(s)}, \vec{\sigma}^{(s')}) =\frac{1}{2}(\mc{N} -\vec{\sigma}^{(s)}\cdot \vec{\sigma}^{(s')})\equiv r$, namely,
\beq
\label{eq:P0}
P_0(r)= \frac{1}{2^\mc{N}}{\mc{N}\choose r}.
\eeq
The main idea of the BID method~\cite{acevedo2025unsupervised} is based on the assumption that if the variables are correlated, they can be described \emph{effectively} by an uncorrelated system living in a lower-dimensional geometrical object. 
Further, it is assumed that its dimension,  $d(r)$, is a smooth function of the distance $r$. Consequently, the following ansatz for the distribution of Hamming distances is proposed 
\beq
\label{eq:fit}
P(r)= \frac{\mc{C}}{2^{d(r)}}{d(r)  \choose r},
\eeq
where $\mc{C}$ is a normalization constant. 
It was empirically found in Ref.~\cite{acevedo2025unsupervised} that the model~\eqref{eq:fit} can capture correctly the observed distributions of distances upon performing a first-order Taylor expansion $d(r)\approx d_0 + d_1 r$, where the BID is defined as the zeroth-order coefficient, i.e.,  $d_\mathrm{BID} \equiv d_0$. 
The parameters $d_0$ and $d_1$ can be determined via standard inference or optimization algorithms. Here we minimize the Kullback-Leibler divergence between the empirical distribution of Hamming distances and the model in Eq.~\eqref{eq:fit}; see Ref.~\cite{acevedo2025unsupervised} and the Supplemental Material~\cite{suppMat}. 
Note also that~\eqref{eq:fit} contains the exact free solution~\eqref{eq:P0} with $\mc{C}=1$, and $d(r)=\mc{N}$.
A key aspect of the BID estimator is that, 
it has been numerically shown to scale linearly with system size for short-range correlated spin systems, for which the number of degrees of freedom scales with a volume law.
This is a far from trivial property in high-dimensional inference, due to the so-called curse of dimensionality~\cite{bishop2016pattern}, namely the fact that, in order to sample \textit{locally} a manifold one needs a number of samples that grows exponentially with its dimension. 
The BID is a global observable, which does not depend on local-neighborhood statistics, making it suitable to work in high-dimensional systems.

Next, we discuss the models and datasets considered in this work.

\emph{Growth models and binary data}. 
Statistical growth models on the lattice are very important because they describe features of real-life nonequilibrium systems as simple stochastic processes. The elementary variables of such systems are non-negative integer-valued \emph{heights} $h_i$, defined on the sites of a $D$-dimensional lattice of linear size $L$ (thus here $\mc{N}=L^D$). During the growth process, the average height $\overline{h}$ increases linearly in time (hence $\overline{h}$ sets a natural unit of time). The dynamics of the growing interface is therefore better characterized by higher moments such as the standard deviation of the heights, or \emph{surface width}, 
\beq
\label{eq:width}
W(L, t) = \Bigg[\Bigg\la  \frac{1}{L^D} \sum_i \Big( h_i(t) - \overline{h}(t) \Big)^2\Bigg\ra  \Bigg]^{1/2},
\eeq
where $\la \cdot \ra$ denotes an average over random realizations. Starting from an initially flat interface, $W$ obeys the Family-Vicsek scaling relation~\cite{Family_1985, PhysRevLett.52.1669}:
\beq
\label{eq:FV}
W(L, t) \sim L^{\alpha} f(t L^{-z}),
\eeq
where $f(u)\sim u^{\beta}$ for $u\ll 1$, and $f=\mathrm{const}$ for $u\gg 1$.  The growth exponent $\beta$ describes an intermediate regime, where $W\sim t^{\beta}$, with an increasing correlation length $\xi_{\parallel}\sim t^{1/z}$ ($z=\alpha/\beta$ is the dynamical exponent). The roughness exponent $\alpha$ characterizes a later saturation regime in which the correlation length has exceeded the size of the system and $W_\mathrm{sat}\sim L^{\alpha}$. Galilean invariance implies the relation $\alpha + z =2$. These exponents do not depend on microscopic details and, therefore, allow for a classification of growth processes into nonequilibrium universality classes~\cite{RevModPhys.76.663, Barabasi_Stanley_1995}. 
A finer classification into surface growth universality classes can be carried out by considering the scaling behavior of the \emph{local} width function $w(l,t)$ [defined in analogy to Eq.~\eqref{eq:width} but with spatial averages over subsystems of linear size $l$] or height-height correlation functions, which may exhibit \emph{anomalous scaling}, with a generally independent local roughness exponent $\alpha_\mathrm{loc} \neq \alpha$~\cite{PhysRevLett.84.2199}.
For an in-depth discussion of surface growth with standard dynamical scaling, the reader is referred to Ref.~\cite{Barabasi_Stanley_1995}.

In the following, we consider two prototypical lattice models for correlated growth. The first system is the restricted solid-on-solid (RSOS) model~\cite{PhysRevLett.62.2289, kim91},  where the deposition of a unit-size particle at a randomly chosen site occurs only if the height difference with its immediate neighboring sites satisfies the restriction $|\Delta h| \le n$, for some integer $n\ge 1$ (here we restrict ourselves to $n=1$; see Ref.~\cite{suppMat} for results with $n>1$). This model belongs to the paradigmatic Kardar-Parisi-Zhang (KPZ) universality class, with exponents $(\alpha, \beta, z)_{1D}^\mathrm{KPZ}=(1/2, 1/3, 3/2)$, in $D=1$. The second system is a model for random deposition with surface diffusion (RDSD)~\cite{Family_1986}, where a unit-size particle is dropped at a randomly chosen site, but it can then move to a column of lower height within a vicinity of the chosen site (here we consider only nearest-neighbor diffusion). The latter model belongs to the Edwards-Wilkinson (EW) universality class,  with exponents $(\alpha, \beta, z)_{1D}^\mathrm{EW}=(1/2, 1/4, 2)$, in $D=1$.

In order to study the role of dimensionality, we also consider the RSOS model in $D=2$. Earlier numerical studies of the two-dimensional (2D) RSOS model~\cite{PhysRevLett.62.2289, kim91} had conjectured rational exponents, $(\alpha, \beta, z)_{2D}=(2/5, 1/4, 8/5)$. However, more recent high-precision simulations~\cite{PhysRevE.92.010101, PhysRevE.94.022107} found irrational values that, although close to the aforementioned rational numbers, rule them out.  Since the goal of the present work is not a high-precision estimation of the scaling exponents, we shall take the rational numbers above as reference.

We performed the Monte Carlo simulations of the models discussed above for different linear sizes $L$, up to a final time $t_f=10^4$, for the $n=1$ RSOS model ($D=1, 2$), and $t_f=2\times 10^4$, for the RDSD model ($D=1$). All simulations start from an empty lattice, and periodic boundary conditions are imposed. For independent random realizations, we save snapshots of the growing interface at different times as datasets: $\mathcal{H}(t) = \{\vec{h}^{(s)}(t)\}_{s=1}^{N_r}$, where $\vec{h}^{(s)}(t)=\big(h_1^{(s)}, h_2^{(s)}, \dots, h_{L^D}^{(s)}\big)_t$ denotes the $s$th realization of the surface profile at time $t$. 
While these numerical simulations generate datasets of integer numbers, in order to apply the BID method, we need to convert those data into $\pm 1$ values. Since no general rule exists to carry out this step, one must rely on schemes that take into account specific features of the system under consideration. For growth processes, the mean height sets a natural reference point to effectively binarize the height variables. Thus, we consider the following binarization rule:
\begin{align}
\label{eq:binary}
h^{(s)}_i &\to  -1, \,\ \mathrm{if} \,\ h^{(s)}_i< \overline{h}, \nonumber \\
h^{(s)}_i &\to +1,  \,\ \mathrm{if} \,\ h^{(s)}_i \ge \overline{h},
\end{align}
where $\overline{h}$ is the mean height at the given time (explicit time dependence has been omitted for convenience).
From a physical perspective, the above transformation closely resembles a ``coarse-grained'' description (\emph{\`a la} Migdal-Kadanoff), in the space of height fluctuations.  

As mentioned above, the choice of the binarization threshold is, in general, arbitrary. Further sensible choices include other measures of central tendency (e.g., the median or the mode) of the empirical height distribution. For the systems under study, these latter measures turn out to be very similar---if not identical---to the mean height, and hence the results presented below are robust with respect to the choice of binarization threshold; see Ref.~\cite{suppMat} for more details on this aspect. 

We note that, in Eq.~\eqref{eq:binary}, we have arbitrarily included the case $h_i^{(s)}=\overline{h}$ to the value $+1$. This naturally induces an imbalance in the number of $-1$'s and $+1$'s, which becomes more severe for datasets with larger $L$ or $D$, due to simple concentration arguments. This can be avoided by a simple modification of the rule in Eq.~\eqref{eq:binary}, in which one sets $h_i^{(s)}$ to either $\pm 1$, at random and with equal probability, when $h_i^{(s)}=\overline{h}$. We have found similar results using either of the two schemes; see Ref.~\cite{suppMat}. Note that, for continuous-variable systems, this consideration becomes irrelevant and one can simply binarize the data with the sign function: $\mathrm{sgn}(\delta h) = -1$ if $\delta h<0$, $\mathrm{sgn}(\delta h) = +1$ if $\delta h>0$, with $\delta h:=h_i^{(s)}-\overline{h}$.

The central question to be addressed below is whether, upon the drastic form of data compression realized in Eq.~\eqref{eq:binary}, the essential correlations are still present, and, in fact, govern the structure of the binarized data. 

\begin{figure}[bt!]
	\centering
	\includegraphics[width=\columnwidth]{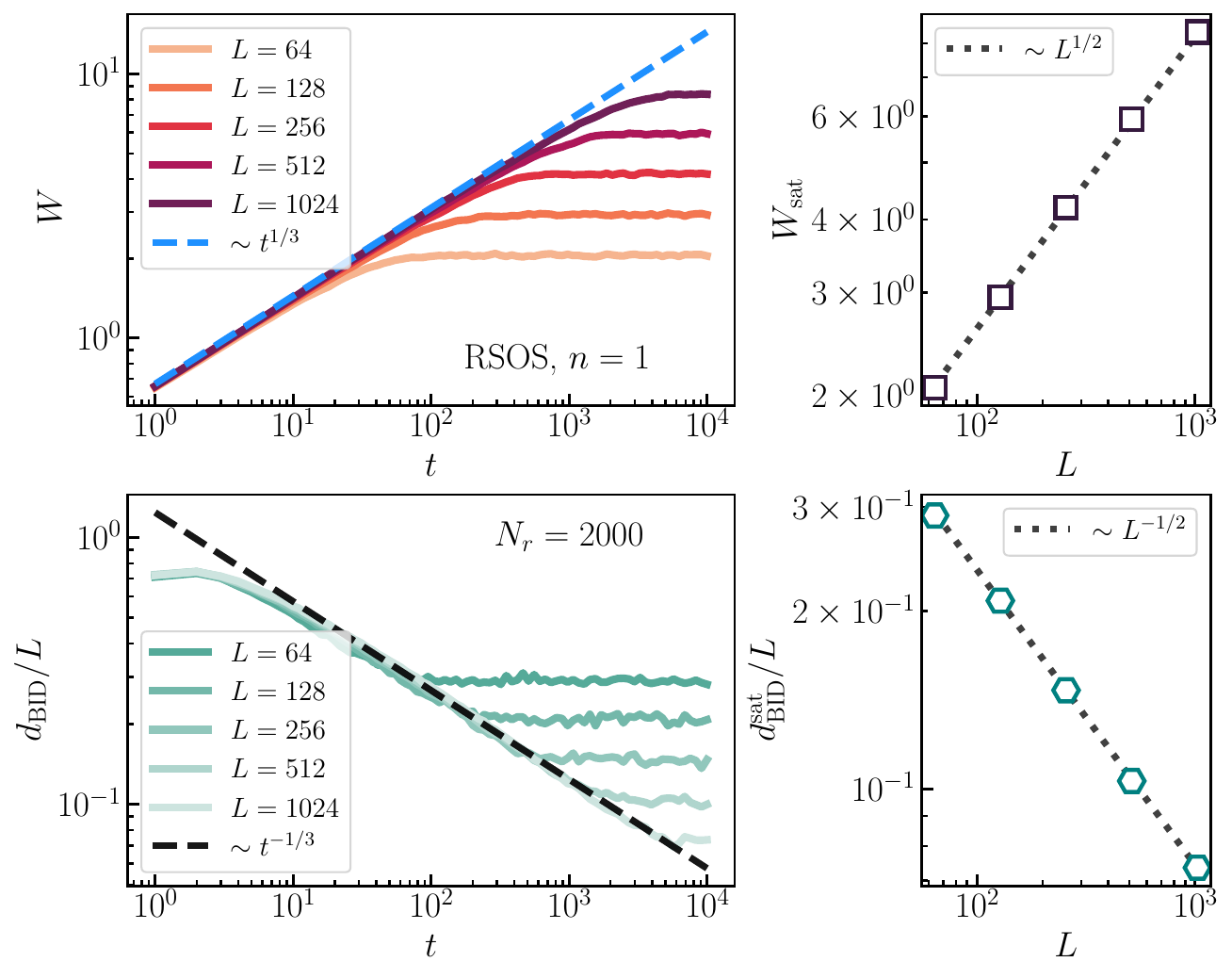}
	\caption{
	Numerical results for the RSOS model ($n=1$) in $D=1$. The upper panels show the scaling behavior of the width  $W$ for different system sizes $L$, in both the intermediate dynamical regime (left) and the stationary regime (right). The lower panels show the corresponding plots for $d_\mathrm{BID}/L$, which also follows an intermediate \emph{decaying} regime as correlations build up in the system (left), and a saturation regime (right). The data shown are computed on datasets with $2000$ samples, at $84$ observation times, which are logarithmically spaced from 1 to $10^4$ (in units of $\overline{h}$). The saturation values are estimated as the average over an \emph{ad hoc} time window for each system size. The dashed and dotted curves are included for visual reference and not obtained from fitting the data.}
	\label{fig:RSOS_N1}
\end{figure}

\begin{figure}[bt!]
	\centering
	\includegraphics[width=\columnwidth]{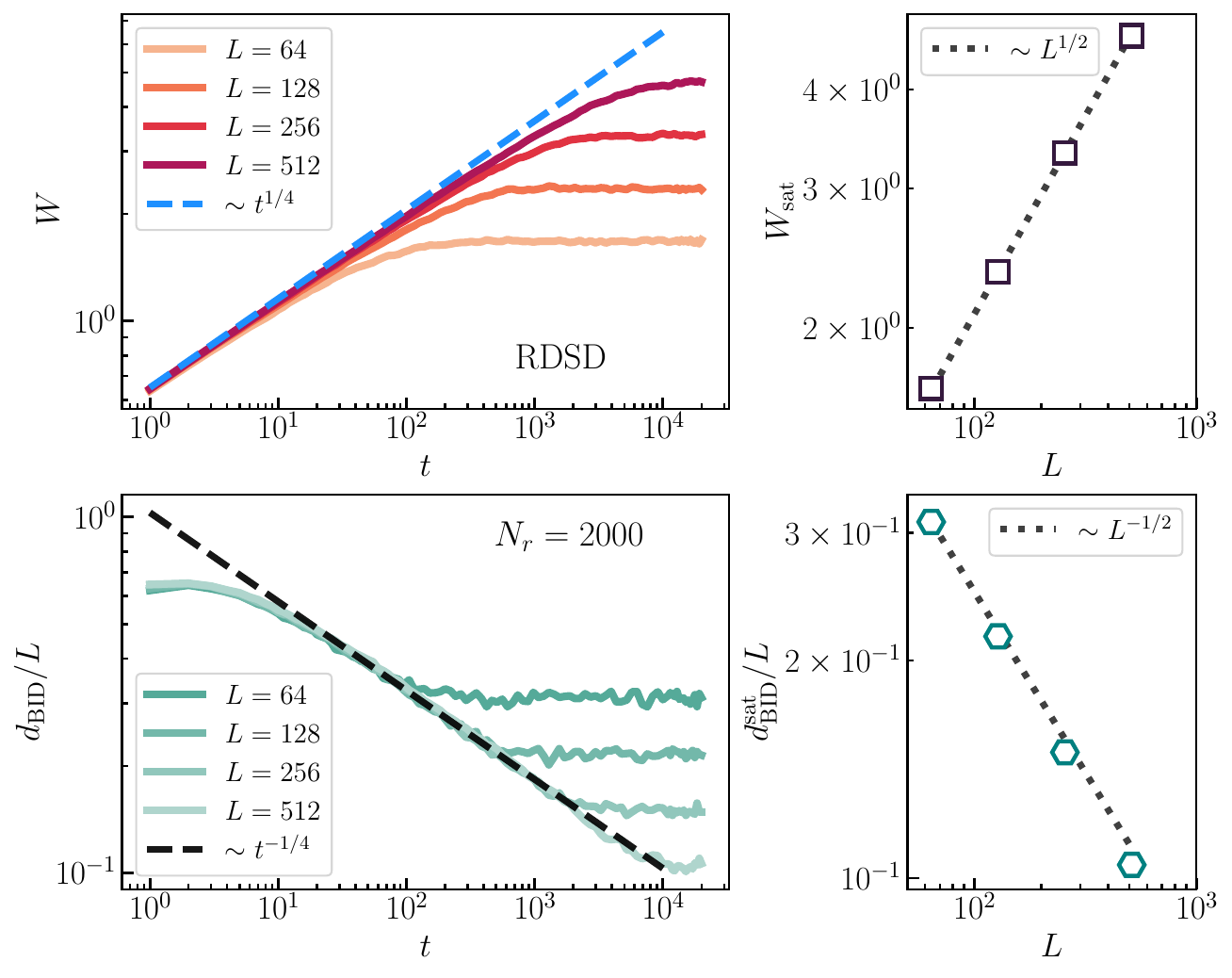}
	\caption{
	Same as in Fig.~\ref{fig:RSOS_N1}, but for the RDSD model in $D=1$.  The data shown are computed on datasets with $2000$ samples, at $100$ observation times, which are logarithmically spaced from 1 to $2 \times 10^4$ (in units of $\overline{h}$). 
	}
	\label{fig:RDSD}
\end{figure}

\begin{figure}[bt!]
	\centering
	\includegraphics[width=\columnwidth]{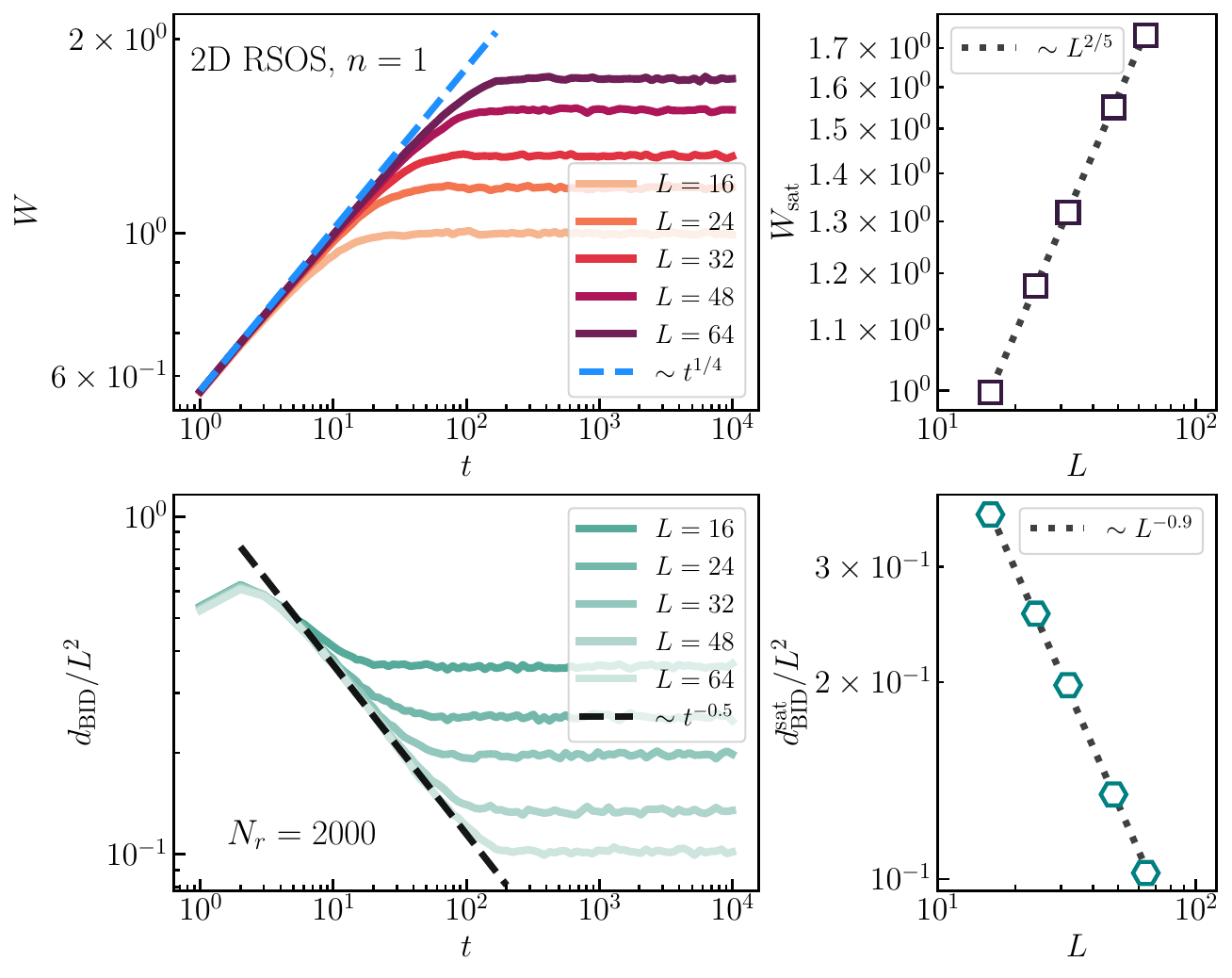}
	\caption{
	Same as in Fig.~\ref{fig:RSOS_N1}, but for the RSOS model ($n=1$) in $D=2$.  Note that here $d_\mathrm{BID}$ is normalized by $L^2$.
	}
	\label{fig:2DRSOS}
\end{figure}



\emph{Scaling of $d_\mathrm{BID}/L^D$}. Shown in Figs.~\ref{fig:RSOS_N1}, \ref{fig:RDSD}, and \ref{fig:2DRSOS} are the numerical results for the temporal behavior of the width function $W$ and $d_\mathrm{BID}/L^D$,  for the RSOS and RDSD models in $D=1$, and the RSOS model in $D=2$, respectively.
These results have been computed on datasets of $N_r=2000$ independent random realizations, at different evolution times. 
While $W$ shows the known behavior reported above, both in the growing and saturation regimes, our main result is the observation that $d_\mathrm{BID}/L^D$ of the binarized data also scales with pristine power laws in the two aforementioned regimes. At intermediate timescales, $d_\mathrm{BID}/L^D$ decays as $\sim t^{-\beta'}$. This dynamical decay captures the expected behavior of the BID, since, physically, there is a correlation length that grows in time ($\xi_\parallel \sim t^{1/z}$), implying that the variables become dynamically correlated over an ever increasing length scale. After a timescale, which roughly coincides with the saturation time of $W$, $d_\mathrm{BID}/L^D$ also features a saturation regime in which it follows a system-size dependence of the form $\sim L^{-\alpha'}$. Remarkably, for $D=1$, our numerical results show a clear direct relationship between the scaling exponents of $W$ and those of $d_\mathrm{BID}/L$, namely, $\alpha\approx \alpha'$ and $\beta \approx \beta'$ (Figs.~\ref{fig:RSOS_N1} and \ref{fig:RDSD}). For $D=2$, $d_\mathrm{BID}/L^2$ also exhibits dynamical scaling. However, our results indicate that $\alpha\approx \frac{1}{2}\alpha'$ and $\beta \approx \frac{1}{2} \beta'$ (Fig.~\ref{fig:2DRSOS}), with a discrepancy in the saturation exponent which is harder to resolve. Further, we note that we have found a higher sensibility of our results to the choice of binarization for $D=2$ than $D=1$~\cite{suppMat}.

\begin{figure}[bt!]
	\centering
	\includegraphics[width=\columnwidth]{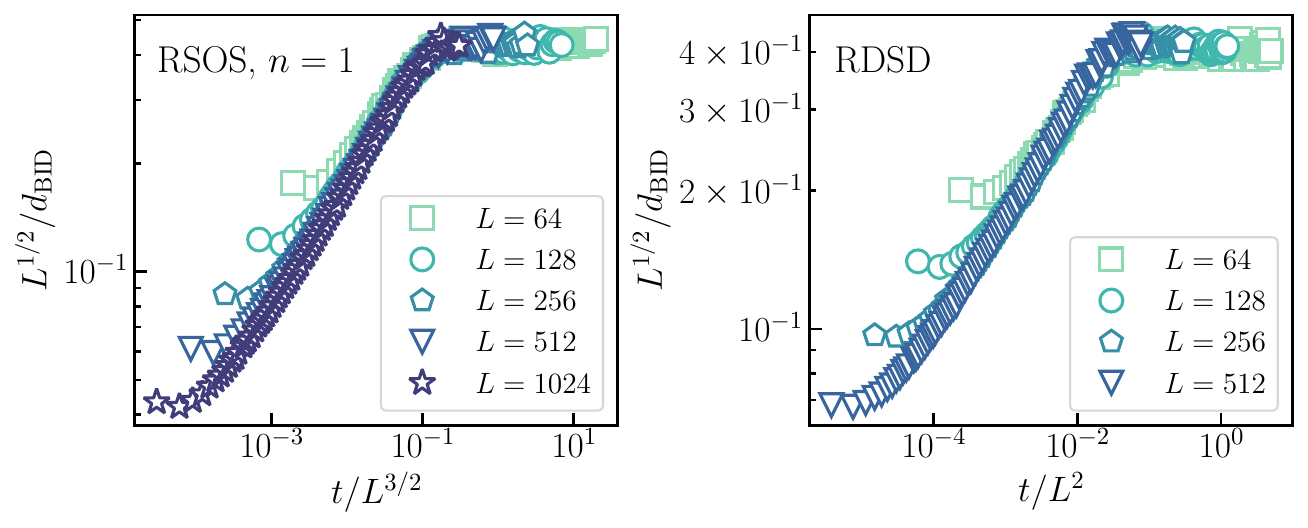}
	\caption{
	Collapse of the data displayed in Figs.~\ref{fig:RSOS_N1} and \ref{fig:RDSD}, for the RSOS and RDSD models in $D=1$, respectively. The axes have been rescaled according to the Family-Vicsek scaling relation in Eq.~\eqref{eq:FV_BID}.
	}
	\label{fig:collapse}
\end{figure}

Motivated by these numerical observations, we expect $d_\mathrm{BID}/L^D$ to obey a Family-Vicsek scaling relation: 
\beq
\label{eq:FV_BID}
\frac{1}{d_\mathrm{BID}} \sim L^{\alpha'- D} g(t L^{-z'}),
\eeq
with $g$ behaving in the same way as $f$ in Eq.~\eqref{eq:FV}, and the extra factor $L^{-D}$ being due to normalization. We have tested this hypothesis by numerically collapsing our data, as shown in Fig.~\ref{fig:collapse}. Since, the numerical results in Figs.~\ref{fig:RSOS_N1} and \ref{fig:RDSD} clearly show a close similarity between the scaling exponents of $W$ and $d_\mathrm{BID}/L$, in our data collapse analysis we have simply used the known scaling exponents of the corresponding universality class. We observe that apart from a transient regime at early times, the data for all system sizes do indeed collapse into a single universal curve. The early nonscaling regime is in contrast to what is observed for $W$, indicating a greater sensibility to the initial condition upon binarizing the data with Eq.~\eqref{eq:binary} (this can also be observed in Figs.~\ref{fig:RSOS_N1}--\ref{fig:2DRSOS}). This is, however, only a transient effect~\cite{suppMat}. We have also checked the scaling hypothesis in Eq.~\eqref{eq:FV_BID} for the 2D RSOS system, yielding a good data collapse as well~\cite{suppMat}.

%
\emph{Discussion and outlook}. We have studied the BID of nonequilibrium data obtained upon binarizing correlated growth dynamics, and shown numerically that it exhibits Family-Vicsek scaling with universal exponents.
Importantly, our methodology allows one to extract scaling exponents in an \emph{unsupervised} manner, and therefore, it is particularly relevant to study other systems with unknown observables to describe their scaling properties.
Providing a theoretical description of our observations remains a central goal for future studies, with the potential to reveal unexplored aspects of how the constraints imposed by the physics of a system ultimately dictate the structure of its data. 
In particular, our results for the 2D system suggest a nontrivial relation between the scaling exponents of $W$ and $d_\mathrm{BID}/L^D$, namely, $\alpha=c \alpha'$ and $\beta=c \beta'$, with $c\neq1$. The exact relationship between $c$ and the spatial dimension $D$ is a key aspect to be elucidated.
A possible route to address these questions is to develop a framework that connects the distribution of Hamming distances (or overlaps) in Eq.~\eqref{eq:fit}, with the order parameters in spin glass theory~\cite{PhysRevLett.50.1946, mezard1987spin}, potentially allowing one to express the BID as a function of the physical parameters of the system.

Our methodology can be easily applied to describe continuous-variable surface growth processes. Further, we expect that a direct application of the BID to intrinsically binary growth processes yields equally informative results due to its very formulation. 
In addition, the results presented here can guide possible applications in other fields. 
For example, in machine learning, studying the BID of a network's weights or activations could shed light on how correlations are processed throughout the network. 
In this context, it would be particularly intriguing to see how the BID behaves around the underparametrized to overparametrized transition~\cite{Biroli-jamming}, or if the BID can capture the scaling exponents in models exhibiting ``neural scaling laws''~\cite{doi:10.1073/pnas.2311878121, hestness2017deep}, in which the loss function behaves as a power law in terms of certain factors, such as the number of parameters or the size of the training set. 

Further interesting directions include exploring possible relations with other compression-based approaches to study nonequilibrium systems, based on exact coarse-graining procedures~\cite{PhysRevLett.125.110601}, or information-theoretic measures of lossless compression of bit strings~\cite{PhysRevX.9.011031, PhysRevE.103.062141}.
Finally, the BID could also be used to characterize nonequilibrium quantum many-body dynamics in systems featuring quantum-fluctuating interfaces or strings~\cite{ krinitsin2025roughening, PhysRevLett.129.120601, PhysRevB.107.024306,  PhysRevB.111.L140305}, including recent experiments with quantum processors~\cite{Gonzalez-Cuadra2025, Cochran2025}.


\emph{Acknowledgments}. We thank Marcello Dalmonte, Alessandro Laio, and Cristiano Muzzi, for insightful discussions and comments on this work. R.V. also acknowledges discussions with Isaac P\'{e}rez Castillo on related projects. R.V. gratefully acknowledges financial support from the ERC Consolidator grant WaveNets. S.A. acknowledges financial support by the region Friuli Venezia Giulia, Project No.~F53C22001770002.

\emph{Data availability}. The datasets used in this work are available on Zenodo in Ref.~\cite{my_zenodo_updated}.

\emph{Code availability}. Routines to compute the BID are implemented in the open-source package DADApy~\cite{GLIELMO2022100589}, and an introductory tutorial can be found in the \href{https://github.com/sissa-data-science/DADApy/blob/main/examples/notebook_hamming.ipynb}{official webpage}.

\bibliography{bid_BIB}

\pagebreak
\widetext
\begin{center}
\textbf{\large Supplemental Material}
\end{center}
\setcounter{equation}{0}
\setcounter{figure}{0}
\setcounter{table}{0}
\setcounter{page}{1}
\makeatletter
\renewcommand{\theequation}{S\arabic{equation}}
\renewcommand{\thefigure}{S\arabic{figure}}
\renewcommand{\bibnumfmt}[1]{[S#1]}
\renewcommand{\citenumfont}[1]{#1}

In this supplementary material, we provide:  (i) further details about the BID method; (ii) results for the RSOS model with higher constraint value $n>1$, including the limit case $n\to \infty$, in which the system reduces to the random deposition model; and (iii) a comparison with a different binarization scheme than the one discussed in the main text.

\section{Further details about the BID method}

Here we discuss more details about the BID method. However, for a more thorough discussion, we refer the reader to Ref.~\cite{acevedo2025unsupervised}. 

As specified in the main text, in practice, one gets an estimate of the BID by fitting the model defined by Eq.~(2) of the main text, with $d(r)\approx d_0 +d_1 r$, to the empirical distribution of Hamming distances computed on the binary dataset of interest. Following Ref.~\cite{acevedo2025unsupervised}, we minimize the Kullback-Leibler divergence between the aforementioned model and the empirical distribution, that is, 
\beq
\label{eeq:KL}
D_\mathrm{KL}(P_{emp}||P)= \sum_{r=r_{min}}^{r_{max}} P_{emp}(r) \log \Bigg( \frac{P_{emp}(r)}{P(r)} \Bigg).
\eeq
There are two meta-parameters in this equation: $r_{max}$ sets an upper cut-off scale that allows to constrained the fit to be local, while $r_{min}$ plays the role of a regularizer, removing very short distances, which can be poorly sampled, generating numerical instabilities. More technically, $r_{max}$ is computed as the quantile of order $\alpha_{max}$ of the empirical cumulative distribution, that is, $\alpha_{max} = \sum_{r=0}^{r_{max}} P_{emp}(r) \in [0,1]$ (and likewise for $r_{min}$). In practice, rather than specifying $r_{max}$ and $r_{min}$, we tune $\alpha_{min}$ and $\alpha_{max}$ so as to get a good fit that minimizes $D_\mathrm{KL}$.

The loss function in Eq.~\eqref{eeq:KL} is minimized stochastically by proposing an update $d_0 \to \delta + d_0$, $d_1 \to \delta +d_1$, where $\delta$ is the step-size parameter, and accepting the move if it reduces the value of $D_\mathrm{KL}$.

All the results shown in the main text have been obtained with $\alpha_{min}=0.1$, $\alpha_{max}=0.7$, with a stochastic optimization of $N_\mathrm{steps}=10^5$ and step size $\delta=5\times10^{-3}$. We did not encounter significant differences with sensible variations of these numbers. However, below we also comment on the effect of running the optimization over more steps. 

\begin{figure}[bt!]
	\centering
	\includegraphics[width=0.64\columnwidth]{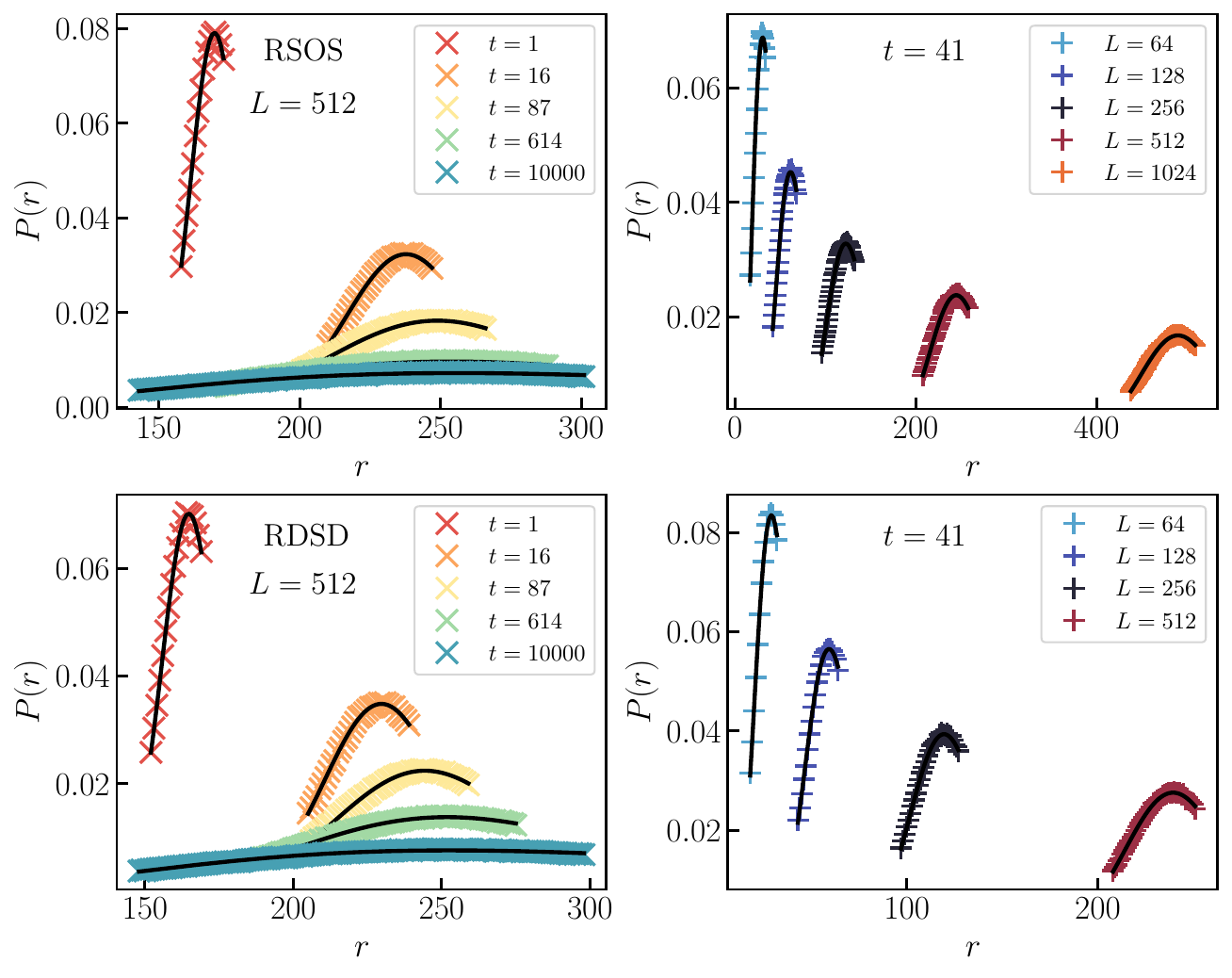}
	\caption{
	  Fits of the distribution of Hamming distances using Eq.~(2) of the main text, for the RSOS, $n=1$ (upper panel) and the RDSD (lower panel) model in $D=1$. The plots on the left show various fits (solid black lines) at different evolution times for a system with $L=512$. The plots on the right show various fits (solid black lines)  for varying system size at $t=41$. The markers denote the empirical distributions of distances over $N_r=2000$ snapshots. All the fits have been done within the window $(\alpha_\mathrm{min}, \alpha_\mathrm{max})=(0.1, 0.7)$, over $N_\mathrm{steps}=10^5$ optimization steps of size $\delta =5\times 10^{-3}$.   }
	\label{fig:fits}
\end{figure}

Examples of the fits used to estimate $\mathrm{BID}$ are shown in Fig.~\ref{fig:fits} for the two 1D systems analyzed in this work, both at fixed size while varying time, and at fixed time while varying size.

\section{RSOS model with $n>1$}

In order to understand the applicability of our observations, in this section we study the BID of the 1D RSOS model, with a larger height constraint value $n$ ($|\Delta h| \le n$).

\subsection{Finite $n$}

\begin{figure}[bt!]
	\centering
	\includegraphics[width=0.56\columnwidth]{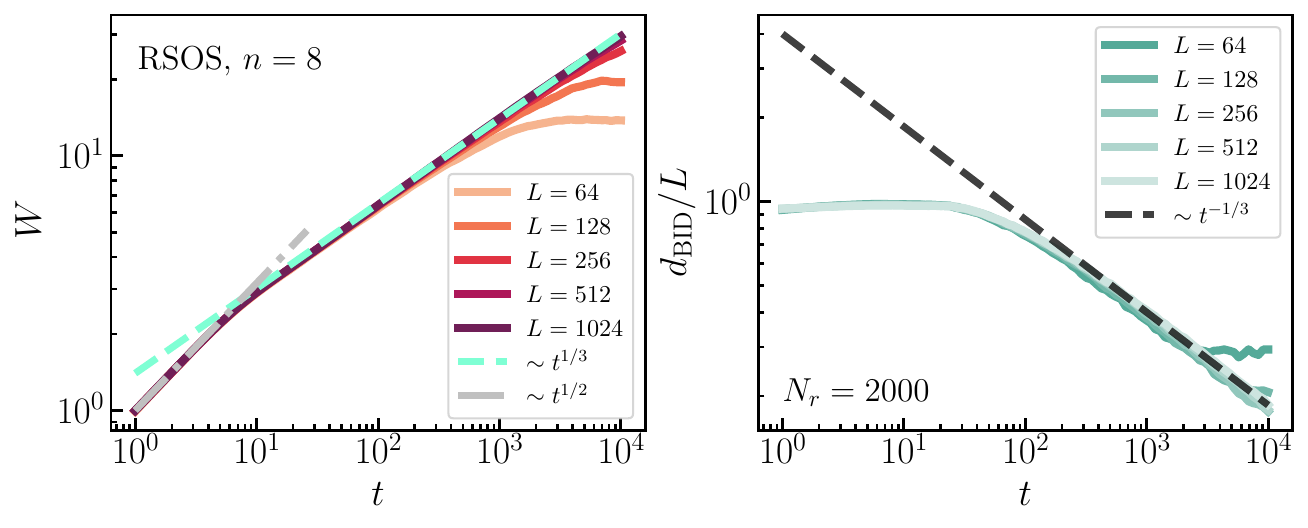}
	\caption{
	Dynamics of the width $W$ (left) and  $d_\mathrm{BID}/L$ (right) for the RSOS model with $n=8$. The effect of a larger integer $n$ is an extended transient regime where the system behaves as random deposition, leading to a smooth crossover from $\sim t^{1/2}$ to $\sim t^{1/3}$ in the growth of $W$. Consequently, $d_\mathrm{BID}/L$ attains a value close to 1 in the random deposition regime, as expected for uncorrelated random data, but it also features a smooth crossover to a decay regime, in which it scales as $\sim t^{-1/3}$, before saturating at  a later time.
	}
	\label{fig:RSOS_n=8}
\end{figure}

The physics of the RSOS model with $n>1$ has been numerically studied~\cite{kim91}. For finite $n$, this leads to an extended initial regime in which the local constraint has no seizable effect on the dynamics of the system. In this regime, the deposition of particles into the substrate essentially as an uncorrelated random process, known as \emph{random deposition}, in which $W\sim t^{1/2}$, in all dimensions. This is followed by a smooth crossover to the same asymptotic behavior as for $n=1$. In Fig.~\ref{fig:RSOS_n=8}, we show numerical results illustrating the physics described above. In terms of $d_\mathrm{BID}/L$, we observe a consistent behavior: in the initial random deposition regime $d_\mathrm{BID}/L\approx 1$, as expected for uncorrelated random data. This is also followed by a slow crossover to a decaying regime, which happens with the same asymptotic behavior as for $n=1$, that is,  $d_\mathrm{BID}/L\sim  t^{-1/3}$.

\subsection{$n\to \infty$: Random deposition}

In the limit of infinite height constraint $n\to \infty$, the RSOS model becomes identical to the random deposition model (RD)~\cite{kim91}, the simplest growth process in which no correlations exist among the points of the surface.  In the RD model, the width grows always as $W\sim t^{1/2}$, in any dimension. Since the growth can continue indefinitely, this universality class is characterized by the growth exponent $\beta$ alone.   

\begin{figure}[bt!]
	\centering
	\includegraphics[width=0.64\columnwidth]{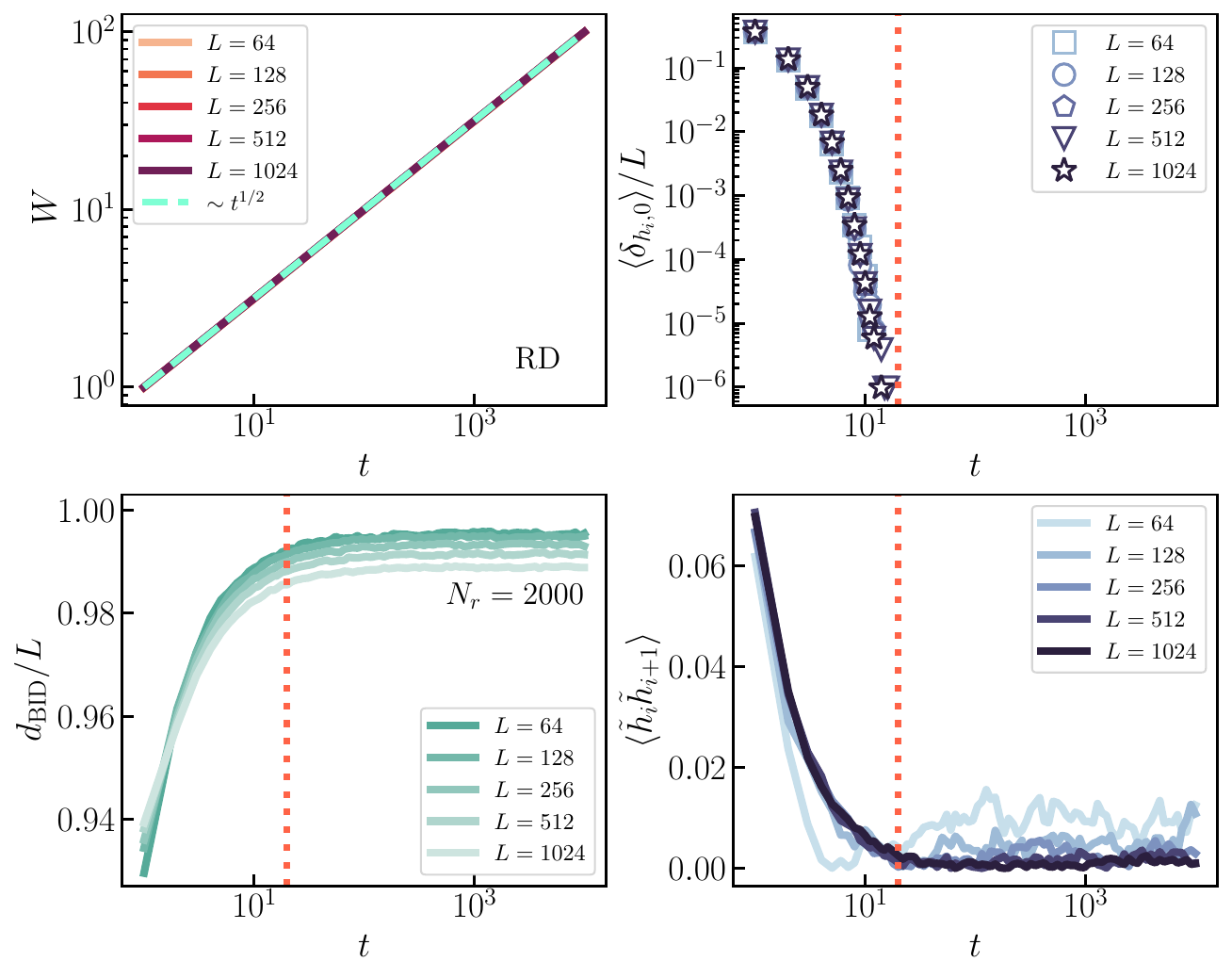}
	\caption{
	Numerical results for the RD model. Evolution of the width $W$ (top, left), the average density of 0's in the original data $\la \delta_{h_i, 0}\ra/L $ (top, right), $d_\mathrm{BID}/L$ (bottom, left), and the spatially averaged nearest-neighbor correlation function $\la \tilde{h}_i \tilde{h}_{i+1} \ra$ (bottom, right)  of the binarized data. A transient ``correlated'' dynamics happens due to the initial condition, in which all height variables are set to 0, and the binarization rule, which leads to an unbalance in the number of $+1$'s and $-1$'s. After the system ``losses'' memory of such initial state, which, according to our numerics, occurs approximately at $t\approx 20$  (dotted orange lines in the different plots), those initial correlations wash away, and the system's behavior becomes truly uncorrelated. Here, all BID fits where performed as global fits, that is, with $(\alpha_\mathrm{min}, \alpha_\mathrm{max})= (0, 1)$.
	}
	\label{fig:RD}
\end{figure}

We have also performed simulations for the RD process. In fact, this case should serve as a perfect benchmark for the BID method: since all height variables are independent, we expect $d_\mathrm{BID}/L\approx 1$. This result is recovered, after a short transient dynamics, in which some correlations are present due to the initial condition (empty substrate) and the binarization rule in Eq.~(5) of the main text; see Fig.~\ref{fig:RD}. 

\begin{figure}[bt!]
	\centering
	\includegraphics[width=0.37\columnwidth]{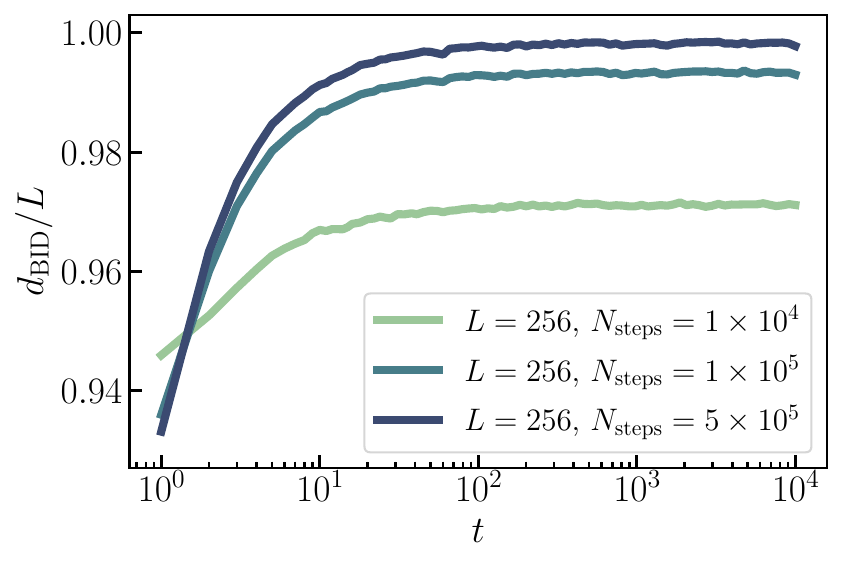}
	\caption{Effect of increasing the number of steps in the optimization loop. For the RD model, we observe a systematic improvement towards the theoretical value $d_\mathrm{BID}/L=1$ for perfectly uncorrelated random binary variables. This is illustrated here for a chain with $256$ sites.
	}
	\label{fig:Nsteps}
\end{figure}

We note that the lower values $d_\mathrm{BID}/L$ upon increasing $L$ is only due to the fact that we have used the same number of optimization steps over all system sizes, leading to a systematic difference in the attained values of the Kullback-Leibler divergence. (For the other models we observed milder differences in the Kullback-Leibler divergence when varying system size, whilst keeping the number of optimization steps fixed.) 
However, upon running our optimization loop longer, these values get closer and closer to 1. This is, for example, illustrated in Fig.~\ref{fig:Nsteps} for a system of size $L=256$. Therefore, if desired, a more fair comparison could be done by self-consistently choosing $N_\mathrm{steps}(L)$ so as to reach (within some tolerance window) a preset value for the Kullback-Leibler divergence.

\section{Different binarization rules}

In this section, we analyze to what extent our results depend on the binarization rule in Eq.~(5) of the main text. In particular, we note that the former rule, due to its own definition, may lead to a finite imbalance between the number of $-1$'s and $+1$'s in the binarized data, since for the case in which $h^{(s)}_i=\overline{h^{(s)}}$, we  have  arbitrarily set $h^{(s)}_i \to +1$.  To remove this bias, here we study the BID of the data binarized according to the following rule:
\begin{align}
\label{eq:binary2}
h^{(s)}_i &\to  -1, \,\ \mathrm{if} \,\ h^{(s)}_i < \overline{h^{(s)}}, \nonumber \\
h^{(s)}_i &\to +1,  \,\ \mathrm{if} \,\ h^{(s)}_i > \overline{h^{(s)}}, \nonumber \\
h^{(s)}_i &\to \mathrm{rand}(-1,+1),  \,\ \mathrm{if} \,\ h^{(s)}_i = \overline{h^{(s)}},
\end{align}
where whenever a given height is equal to the average height, we randomly select either $-1$ or $+1$. 

\begin{figure*}[bt!]
	\centering
	\includegraphics[width=0.81\textwidth]{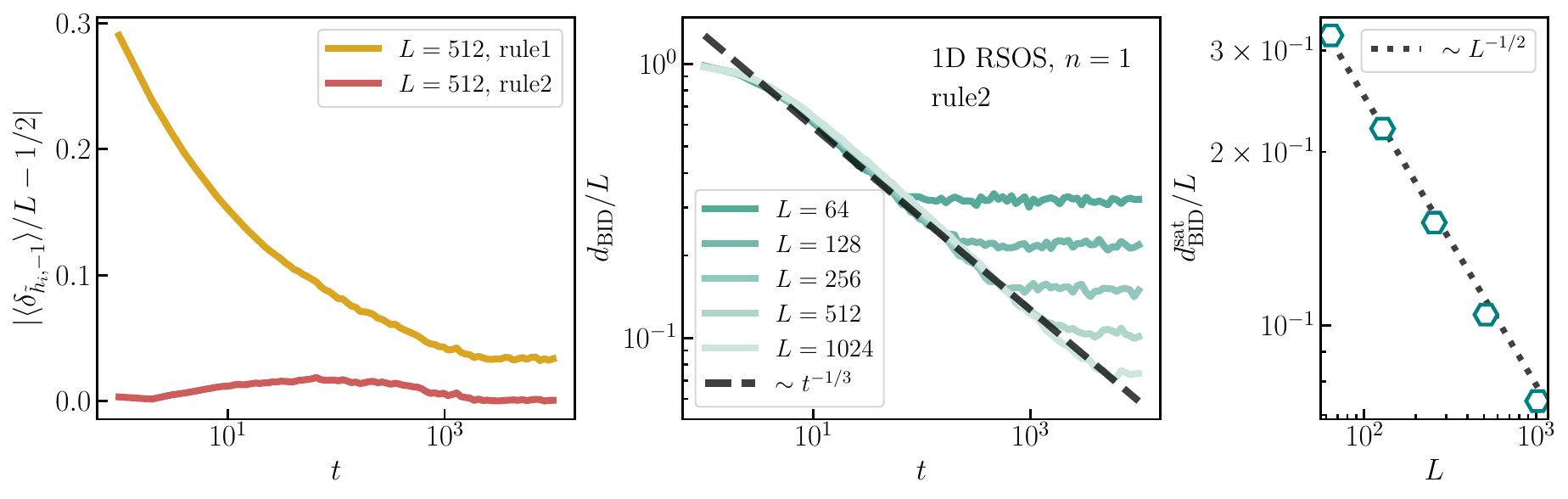}
	\caption{(Left) Absolute deviation from $1/2$ of the average fraction of $-1$'s in the binarized data using rules in  Eq.~(5) of the main text  [``\textsf{rule1}''] and Eq.~\eqref{eq:binary2}  [``\textsf{rule2}''].
	(Middle) $d_\mathrm{BID}/L$ vs.~$t$ for varying $L$, upon having binarized the 1D RSOS ($n=1$) data according to \textsf{rule2} instead of \textsf{rule1} [cf. Fig.~2 of the main text]. The data here are also compatible with a decay as $\sim t^{-1/3}$. (Right) Saturation value of $d_\mathrm{BID}/L$ vs. $L$ when using \textsf{rule2}, exhibiting a compatible behavior with $\sim L^{-1/2}$. 
	}
	\label{fig:rule2}
\end{figure*}

\begin{figure*}[bt!]
	\centering
	\includegraphics[width=0.81\textwidth]{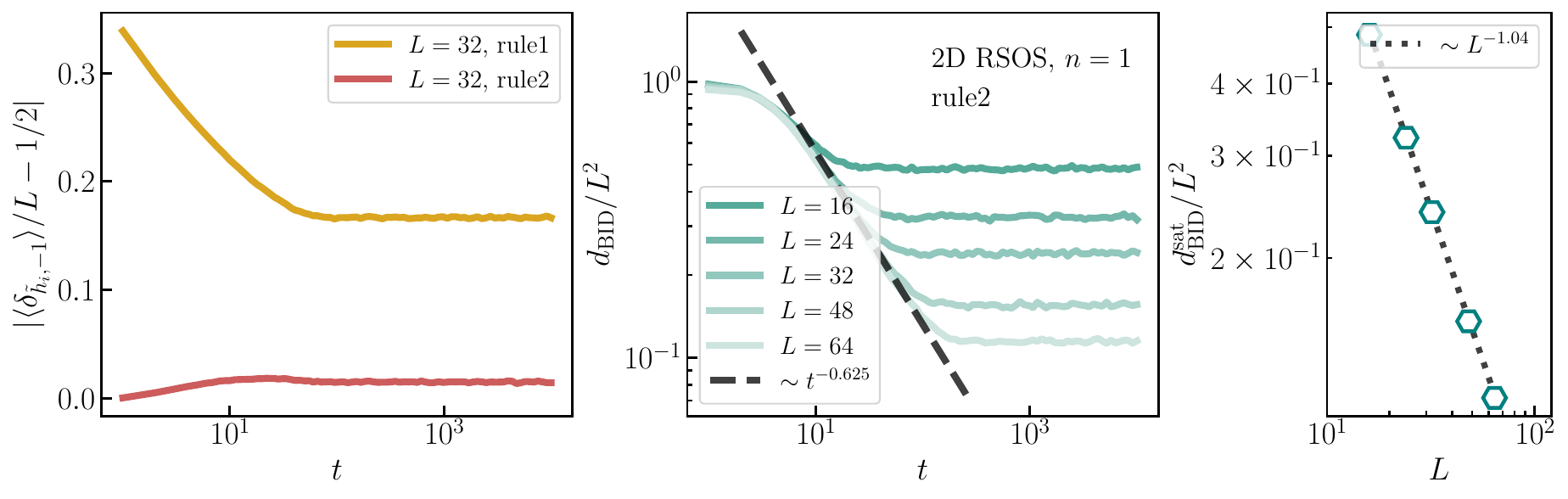}
	\caption{(Left) Absolute deviation from $1/2$ of the average fraction of $-1$'s in the binarized data using rules in  Eq.~(5) of the main text  [``\textsf{rule1}''] and Eq.~\eqref{eq:binary2}  [``\textsf{rule2}''].
	(Middle) $d_\mathrm{BID}/L^2$ vs.~$t$ for varying $L$, upon having binarized the 2D RSOS ($n=1$) data according to \textsf{rule2} instead of \textsf{rule1} [cf. Fig.~4 of the main text]. The data here show a decay compatible with $\sim t^{-0.625}$ (note that, $5/8=0.625$). (Right) Saturation value of $d_\mathrm{BID}/L^2$ vs. $L$ when using \textsf{rule2}, exhibiting a compatible behavior with $\sim L^{-1.04}$. 
	}
	\label{fig:rule2_2D}
\end{figure*}

For the 1D RSOS model with $n=1$, we find that using either of the two binarization rules yields, in fact, similar results. This is shown in Fig.~\ref{fig:rule2}. In particular, we see that $d_\mathrm{BID}/L$ still decays as $\sim t^{-1/3}$, while it saturates as $\sim L^{-1/2}$. In this figure, we also show a comparison between the average fraction of $-1$'s in the binary datasets following the rules in Eq.~(5) of the main text and Eq.~\eqref{eq:binary2}.

For the 2D RSOS model with $n=1$, we find that the choice binarization rule has a more marked effect, as illustrated in Fig.~\ref{fig:rule2_2D}. Although $d_\mathrm{BID}/L$ still exhibits dynamical scaling, it does so with different scaling exponents $\alpha', \beta'$. In particular, our numerical results show that here $d_\mathrm{BID}/L$ decays as $\sim t^{-0.625}$, while it saturates as $\sim L^{-1.04}$. In this case, the relation to the exponents of the width function $W$ is given by $\alpha \approx \frac{2}{5}\alpha'$ and $\beta \approx \frac{2}{5}\beta'$. In this figure, we also show a comparison between the average fraction of $-1$'s in the binary datasets following the rules in Eq.~(5) of the main text and Eq.~\eqref{eq:binary2}.

\begin{figure}[bt!]
	\centering
	\includegraphics[width=0.56\columnwidth]{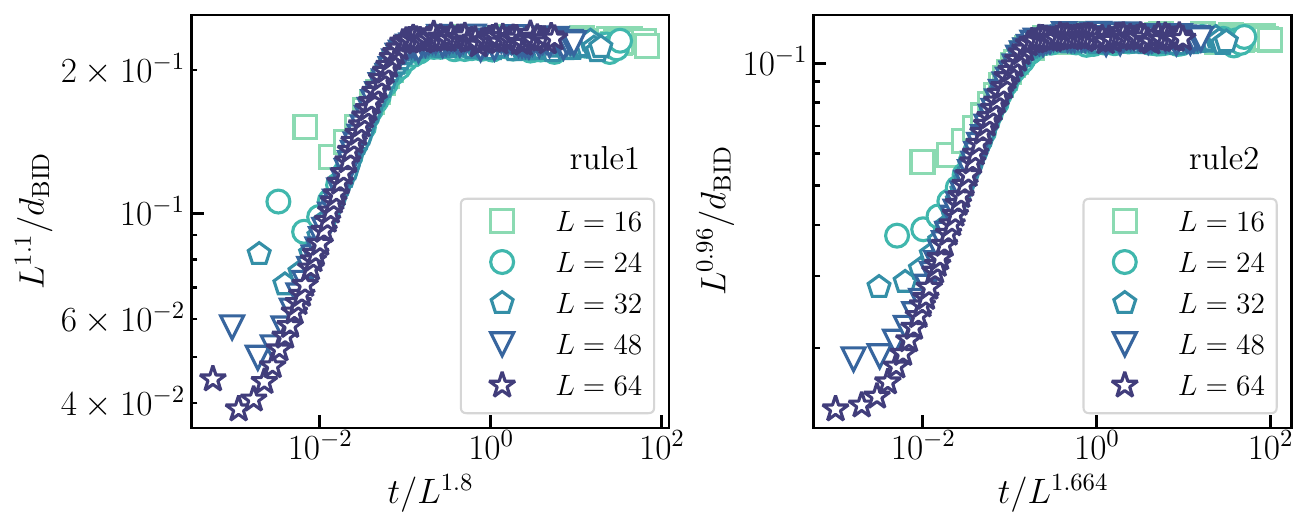}
	\caption{Data collapse of the data of the 2D RSOS ($n=1$) model through the Family-Vicsek scaling relation in Eq.~(6) of the main text, using \textsf{rule1} (left) and \textsf{rule2} (right).
	}
	\label{fig:collapse_2D}
\end{figure}

Finally, in Fig.~\ref{fig:collapse_2D}, we show the data collapse using the Family-Vicsek scaling hypothesis [Eq.~(6) in the main text], for the 2D RSOS model data when using \textsf{rule1} [cf. Fig.~4 of the main text] and \textsf{rule2} [cf. Fig.~\ref{fig:rule2_2D}].

\subsection{Choice of the binarization threshold}

\begin{figure}[bt!]
	\centering
	\includegraphics[width=0.37\columnwidth]{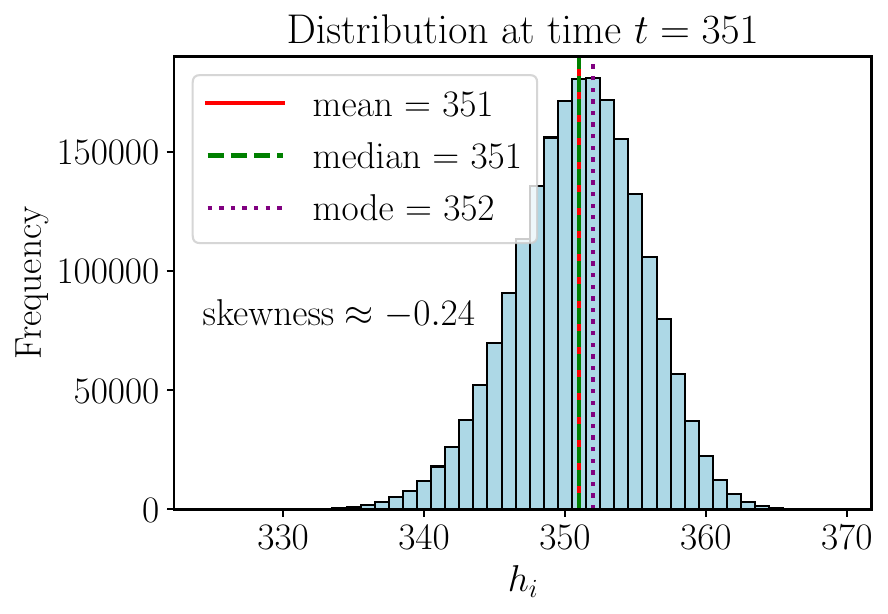}
	\caption{Empirical distribution of the data of the 1D RSOS model with $n=1$ and $L=1024$, at time $t=351$.} 
	\label{fig:height_dist}
\end{figure}

Another important aspect about our binarization scheme is the choice of the threshold used to split the original data into $\pm1$ values. As pointed out in the main text, this choice is arbitrary. The main results in the text were obtained using the sample mean height as the threshold. However, it is pertinent to ask: what happens if one considers other \emph{typical} values of the height distribution to this end?  For the RDSD model, the probability density function (PDF) is asymptotically a Gaussian distribution, and hence, we expect any central tendency measure of the empirical distribution to work as well. However, the question above is particularly relevant when the PDF is not Gaussian, as is the case for the systems in the KPZ universality class.

We have analyzed other sensible choices of the threshold, namely,  the median and mode of the empirical height distribution, and found that, for the models under study, they are actually very close---if not equal---to the mean height in all cases. As said above, this is expected for the RDSD model, but it turns out to be also the case for the RSOS model. As an illustration, in Fig.~\ref{fig:height_dist}, we show the empirical height distribution for the data of the 1D RSOS model with $n=1$ and $L=1024$, at an intermediate time $t=351$, within the scaling regime. Although the distribution is clearly asymmetric, having a skewness of approximately $-0.24$, the mean and median of the sample are the same, while the mode is just one unit above the other two quantities. Similar observations hold for other times and system sizes. Hence, our results remain essentially unchanged regardless of which of these values is used as the threshold. 
We note that the fact that these quantities coincide or are very close to each other, despite the distribution being asymmetric, can indeed occur due to the \emph{discreteness} of the data; see  Ref.~\cite{vonHippel01012005} for an extended discussion on this and related aspects.
The skewness reported above was computed with the Python library \texttt{SciPy}~\cite{2020SciPy-NMeth}, as the Fisher-Pearson coefficient of skewness, namely, $M_3/M_2^{3/2}$, with $M_i$ being the biased \emph{i}th central moment of the sample.

\end{document}